# Prediction of High Temperature Superconductivity in *C*2/*c*-24 Solid Hydrogen

Mehmet Dogan[1,2], Sehoon Oh[1,2], Marvin L. Cohen[1,2,*]

[1] Department of Physics, University of California, Berkeley, CA 94720, USA

[2] Materials Sciences Division, Lawrence Berkeley National Laboratory, Berkeley, CA 94720, USA

[*] To whom correspondence should be addressed: mlcohen@berkeley.edu

## Abstract

Recent experimental developments in hydrogen-rich materials at high pressures have put this class of materials above others in the race toward room temperature superconductivity. As it is the basis of all the materials in this class, the efforts to determine the properties of pure solid hydrogen at high pressures remain intense. Most notably, a recent experimental study of the metallization of hydrogen identified the crystal phase of the solid as the *C*2/*c*-24 molecular phase up to ~425 GPa [P. Loubeyre *et al.*, *Nature* **577**, 631 (2020)]. It is possible that the observed metallization is caused by band structure effects and not a structural phase transition, and the material remains in this crystal phase up to higher pressures [M. Dogan *et al.*, *J. Phys.: Condens. Matter* **33** 03LT01 (2020)]. Therefore it is of crucial importance to determine the superconducting properties of the *C*2/*c*-24 phase. Here, we employ a Wannier function-based dense *k*-point and *q*-point sampling to compute the electron–phonon coupling and superconducting properties of molecular hydrogen in the *C*2/*c*-24 phase. We find that the material has a high superconducting transition temperature of 242 K at 500 GPa. We also find that the transition temperature rapidly increases with pressure in the 400 – 500 GPa range.

Hydrogen was predicted to transition to an atomic crystal and a metal in its solid form in 1935 [1], which was followed by the prediction that it would have a high superconducting transition temperature in 1968 [2]. However, determining the transition to the atomic phase as well as predicting the crystal structure of the molecular phase of solid hydrogen was elusive for several decades until the early 2000s, when a collection of candidates were discovered [3–6]. In the following two decades, through the increasingly sophisticated computational methods and steadily improving experimental studies, the leading candidates emerged as the three molecular phases (*C2/c*-24, *Cmca*-12, *Cmca*-4) and one atomic phase (*I4$_1$/amd*-2) (number after the dash denotes the number of atoms in the unit cell) [7–16]. As the pressure increases, hydrogen may metallize either via the closure of the band gap in a molecular phase or a structural phase transition into a different molecular phase (whose gap is already closed at that pressure) or an atomic phase. Computational studies have not definitively shown which of these scenarios should occur and at which pressure beyond the consensus that several phases become competitive in the 300 – 500 GPa range. The reason is that hydrogen nuclei are too light for the common static nuclei approximation to be accurate, and different sets of phase transitions are predicted when different methodologies of accounting for the quantum nature of the nuclei are used.

Because of the developments in the diamond anvil cell technology, the accessible pressure range gradually reached 400 GPa in the 2000s and 2010s, resulting first in the observation of black hydrogen (a semiconductor with a direct gap below the visible range) around 310 – 320 GPa [17–20], and second in the observation of increased conductivity around 350 – 360 GPa, where solid hydrogen might become semimetallic [21,22]. Recently, an experiment by Loubeyre *et al.* [23] employed infrared (IR) absorption measurements to track the vibron frequency and the direct electronic band gap with pressure up to 425 GPa. In this experiment, it was observed that the IR-active vibron frequency linearly decreases with pressure from 150 GPa to 425 GPa, which indicates that there is likely no phase transition in this pressure range. At the same time, the direct band gap gradually decreases between 360 – 420 GPa, but abruptly drops below the minimum experimentally available value of ~0.1 eV. Although there are important methodological disagreements between different experimental groups [24–26], there is general agreement that some kind of band gap closure occurs in the 425–440 GPa range, as also indicated by the abrupt saturation of IR absorption around 425 GPa observed by Dias *et al.* [21] and similar measurements in the Raman spectra around 440 GPa by Eremets *et al.* [22]. Finally, it is possible that a phase transition to atomic metallic hydrogen occurs around 500 GPa [27,28].

In a previous work [29], we showed that the vibron frequencies of the *C2/c*-24 phase calculated in the anharmonic regime match well with the observations of Loubeyre *et al.* [23] up to 425 GPa. We also demonstrated that it is possible to interpret the observed changes in the direct band gap as a series of pressure induced changes in the band structure without the need to postulate a structural phase transition. Hence, it is possible that the material remains in the *C2/c*-24 phase at higher pressures, potentially up to 500 GPa. Although the atomic and electronic structures of this phase have been reported previously [5,11,29], due to its 24-atom unit cell, its superconducting properties have remained elusive. However, with all the recent developments in hydrogen-rich materials at high pressures that are approaching room temperature superconductivity [30–34], it is even more crucial to understand the superconductivity of this pure hydrogen structure.

Here, we investigate the electronic structure, vibrational properties, electron–phonon coupling and superconducting properties of molecular hydrogen in the *C2/c*-24 phase at 400, 450 and 500 GPa using density functional theory (DFT) calculations in the generalized gradient approximation, anharmonic corrections with the self-consistent phonon approach and a Wannier function-based dense *k*-point and *q*-point sampling (see the **Supplemental Material** for details [35], and references [36–45] therein).

The atomic structure of the *C2/c*-24 phase is shown in **Figure S1**, which consists of van der Waals-bonded layers of molecular hydrogen. The distorted hexagonal unit cell consists of 24 atoms where 6 atoms (3 molecules) lie in 4 inequivalent planes. The Fermi surface and the band structure for 400 and 450 GPa are shown in **Figure S2** and **Figure S3**, respectively. As pressure increases, the material first transitions from a semiconductor to semimetal with a valence band (band 1) and a conduction band (band 2) crossing the Fermi energy (**Figure S2**). As the pressure increases, the Fermi surface consisting of the band 1 and band 2 sheets grows, and other bands start crossing the Fermi energy in the vicinity of the Z-point (**Figure S3**). At 500 GPa, the Fermi surface consists of many electron and hole pockets and with 5 bands crossing the Fermi energy at various points in the Brillouin zone, as shown in **Figure 1**. The density of states at the Fermi energy is 0.002, 0.010 and 0.022 states/eV/atom for 400, 450 and 500 GPa respectively.

We note here that we use the generalized gradient approximation (GGA) which underestimates band gaps. A more accurate treatment of excited states such as the GW approximation is expected to raise band gaps by ~1.5 eV in this pressure range [7,10]. A similar and opposite effect is expected to arise from the treatment of nuclear quantum effects [46–50]. These opposite effects likely happen to cancel out to a large degree, which is likely responsible for the close agreements between the computed and measured direct gap and transition pressure values reported in our previous work [29].

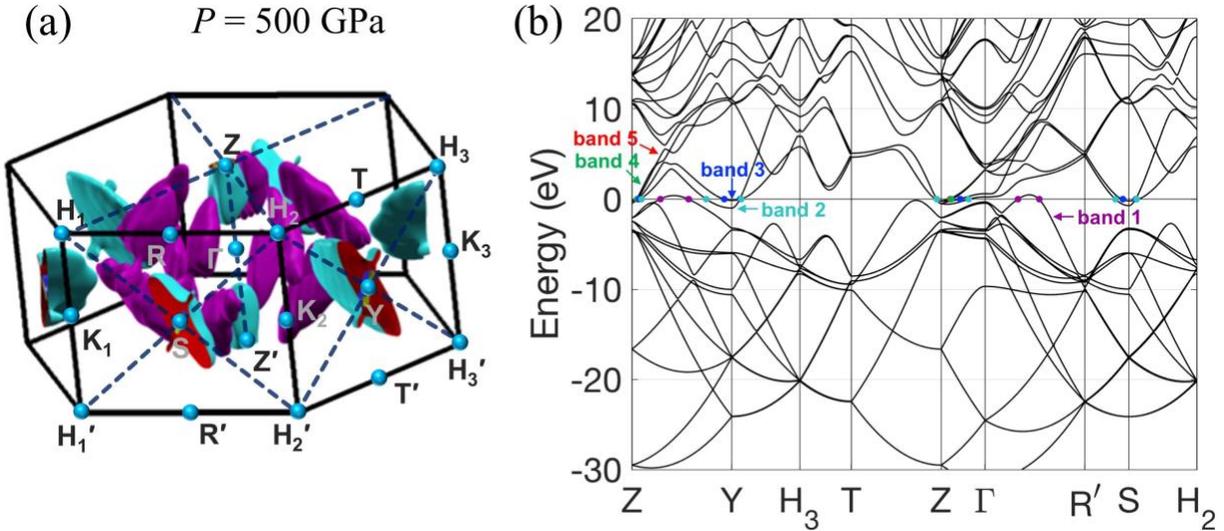

**Figure 1. Fermi surface and band structure at 500 GPa. (a)** The first Brillouin zone and the Fermi surface of the *C2/c*-24 phase of hydrogen at 500 GPa pressure, with high-symmetry points labeled. **(b)** Band structure of the *C2/c*-24 phase of hydrogen at 500 GPa pressure. Energies are relative to the Fermi energy. The bands that cross the Fermi energy are pointed out and the band crossings are labeled with filled circles. The colors of these labels match the outer surface colors of the Fermi surface sheets in (a).

The phonon dispersion relations of the *C2/c*-24 phase at 500 GPa both in the harmonic and anharmonic approximations are presented in **Figure 2**. We observe that the anharmonic corrections increase the lower phonon frequencies, which correspond to collective motion of the atoms within a plane, while they decrease the higher phonon frequencies which correspond to intramolecular vibrations (vibrons). The Eliashberg function ($\alpha^2 F$) and the phonon densities of states (PhDOS) are also presented in **Figure 2**. The $\alpha^2 F$ and PhDOS are qualitatively similar, indicating that the electron–phonon coupling is not provided by specific phonon modes. We also show the electron–phonon coupling parameter defined as $\lambda(\omega) = \int_0^\omega \frac{d\omega'}{\omega'} \alpha^2 F(\omega')$ in **Figure 2**. Although the overall shape of $\lambda(\omega)$ in the harmonic and anharmonic cases are similar, the anharmonic values are smaller because the lower phonon frequencies are pushed up and the integral includes the frequency in the denominator. The electron–phonon coupling constant $\lambda$ corresponds to $\lambda(\infty)$ and is equal to 1.80 and 1.53 for the harmonic and anharmonic cases, respectively. The analogous plots for 400 and 450 GPa are presented in **Figure S4** and **Figure S5**, respectively, and can be interpreted in the same way.

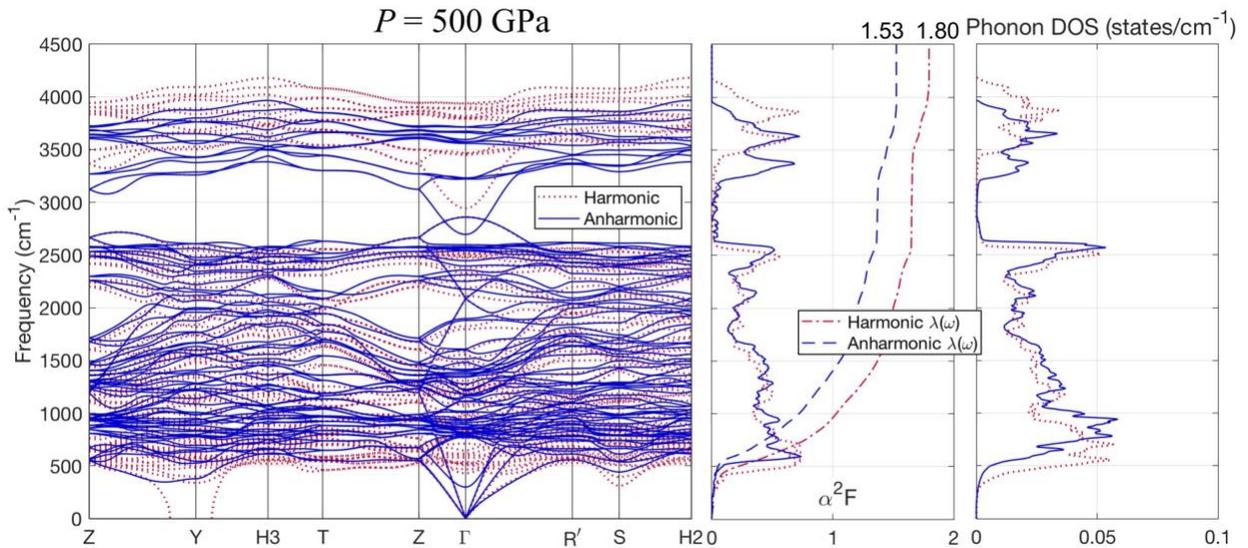

**Figure 2. Phonons and electron–phonon coupling at 500 GPa.** The phonon dispersion relations for the *C2/c*-24 phase of hydrogen at 500 GPa pressure (left panel). The harmonic and anharmonic calculations are shown by red dashed lines and blue solid lines, respectively. The Eliashberg function $\alpha^2 F$ and the

electron–phonon coupling parameter $\lambda(\omega)$ (middle panel), and the phonon densities of states (right panel).

The resulting values of the superconducting transition temperatures ($T_c$) using the Allen–Dynes formula [51] ($\mu^* = 0.1$) are reported in **Table 1**. We see that the $T_c$ values for harmonic and anharmonic cases are close despite the fact that $\lambda$ reduces significantly when anharmonic effects are included. This is due to the fact that $\omega_{log}$ values for the anharmonic calculations are significantly higher, and the two effects mostly cancel out. We note that we use a standard value of 0.1 for the semi-empirical Coulomb pseudopotential to arrive at these $T_c$ values. First-principles estimations of this parameter vary between 0.08 and 0.16 for hydrogen in the *Cmca*-4 phase [52]. For example, using the ends of this interval would yield an Allen–Dynes $T_c$ between 147 K and 201 K at 500 GPa.

**Table 1. Electron–phonon coupling constant and $T_c$.** The electron–phonon coupling constant $\lambda$ and the superconducting transition temperature $T_c$ using both the Allen–Dynes formula and the isotropic Eliashberg theory are shown in the harmonic and anharmonic cases for 400, 450 and 500 GPa. The Coulomb pseudopotential $\mu^*$ is set to 0.1 in all cases.

| *P* (GPa) | $\lambda$ | | $T_c$ (Allen–Dynes) (K) | | $T_c$ (Eliashberg) (K) | |
|---|---|---|---|---|---|---|
| | harmonic | anharmonic | harmonic | anharmonic | harmonic | anharmonic |
| **400** | 0.43 | 0.32 | 8.5 | 1.6 | 8.5 | 1.6 |
| **450** | 0.95 | 0.82 | 79 | 83 | 100 | 94 |
| **500** | 1.80 | 1.53 | 186 | 190 | 245 | 242 |

It has been shown that in the case of high anisotropy in electron–phonon coupling in the $k$-space, superconducting properties are not captured accurately by the isotropic Eliashberg theory [45,53]. However, the computational cost of anisotropic Eliashberg calculations is very high, and isotropic calculations are preferred when possible. To check the degree of anisotropy in the $k$-space, we plot the band and $k$-point resolved electron–phonon coupling ($\lambda_{n,\mathbf{k}}$) in **Figure 3(a)** for 500 GPa in the anharmonic approximation. We find that $\lambda_{n,\mathbf{k}}$ is mostly isotropic, with some anisotropy in the hole pockets (band 2). The analogous plot for the harmonic case is presented in **Figure S6**. We then proceed to the isotropic Eliashberg calculations to determine leading edge of the superconducting gap ($\Delta_0$) vs. temperature, which is shown in **Figure 3(b)** for all the computed pressures in the harmonic and anharmonic approximations. The $T_c$ values obtained from these calculations are also reported in **Table 1**. The rapid increase of $T_c$ with pressure can largely be explained by the rapid increase of the electronic density of states at the Fermi level.

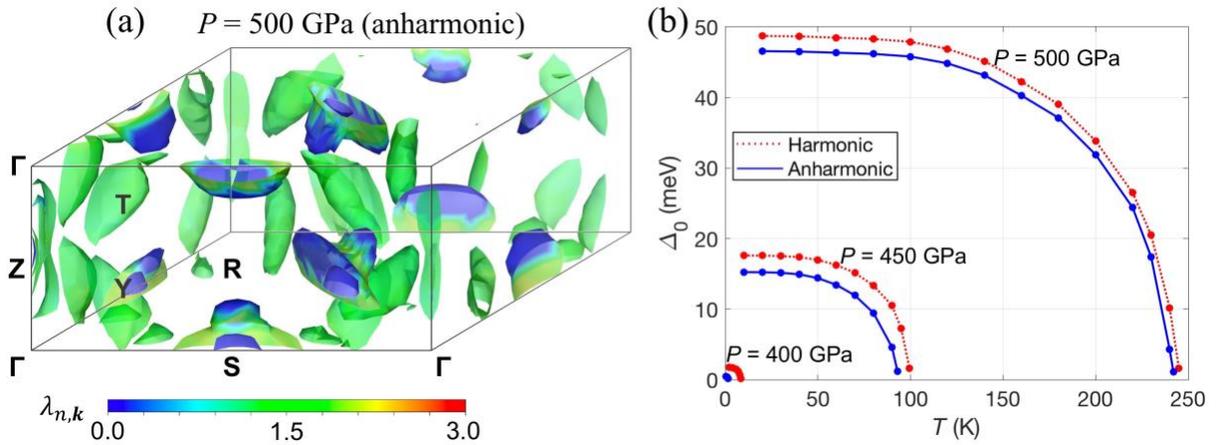

**Figure 3. Electron–phonon coupling and superconducting gap. (a)** Band and $k$-point resolved electron–phonon coupling for 500 GPa for the *C2/c*-24 phase of hydrogen in the anharmonic approximation shown in the reciprocal space. **(b)** Leading edge of the superconducting gap *vs.* temperature for 400, 450 and 500 GPa in the harmonic and anharmonic approximations.

We find that the expected $T_c$ at the 500 GPa (242 K) is very high and comparable to the predicted $T_c$ of the *Cmca*-4 phase at 450 GPa (258 K) [54]. In contrast to the previous work on the *Cmca*-4 phase [54], anharmonicity does not greatly enhance superconductivity in the *C2/c*-24 but instead very slightly diminishes it. We note that the inclusion of nuclear quantum effects and a better treatment of electron self-energy effects (such as the GW method) would change these reported values, especially if these effects occur in substantially nonuniform ways in the reciprocal space. However, we do not think this is likely, since the electronic and vibrational structure we base these calculations on are consistent with the experimental measurements [29]. It is more likely that the opposite effects of the quantum behavior of the nuclei and electron self-energy are largely uniform and similar in size, and the overall qualitative picture, *i.e.*, that $T_c$ rapidly increases in the 400–500 GPa range from <10 K to >200 K, is a robust result. We also note that recently a parameter based on the electron localization function (ELF) was proposed as highly correlated with $T_c$ in hydrides [55]. This parameter, called the networking value ($\phi$), corresponds to the largest value of ELF for which the isosurface of ELF creates a connected network in all three directions. For the *C2/c*-24 phase, we find $\phi$ to be 0.271, 0.275 and 0.276 at 400, 450 and 500 GPa, respectively. In **Figure S1**, we present the plot of the ELF isosurface at the value $\phi = 0.276$ for 500 GPa. Finally, in order to provide another theoretical datapoint for experimental studies that may measure optical properties of high-pressure hydrogen samples, we present the dielectric function and reflectivity of the *C2/c*-24 phase 500 GPa calculated in the random phase approximation (RPA) in **Figure S7**.

In summary, we have found that the predicted superconducting transition temperatures for the *C2/c*-24 phase at 400, 450 and 500 GPa are 1.6, 94 and 242 K, respectively. The rapid rise of $T_c$ with pressure can be understood by the transition of the material from semimetal to metal in that pressure range, as we detailed in our previous work [29]. We find that anharmonic corrections significantly change the electron–phonon coupling, but their overall impact on the superconducting properties and $T_c$ is small. Given that the *C2/c*-24 phase may be stable up to the transition to the atomic phase at a higher pressure around 500 GPa, we hope that our work will motivate experimental researchers to investigate the superconducting properties of hydrogen in the 400 – 500 GPa range. Such experiments could bring us closer to resolving one of the outstanding questions in physics, which is to understand the high-pressure properties of the simplest and the most common material in nature.

**Acknowledgements**

This work was supported primarily by the Director, Office of Science, Office of Basic Energy Sciences, Materials Sciences and Engineering Division, of the U.S. Department of Energy under contract No. DE-AC02-05-CH11231, within the Theory of Materials program (KC2301), which supported the structure optimization and calculation of vibrational properties. Further support was provided by the NSF Grant No. DMR-1926004 which supported the determination of electron–phonon interactions. Computational resources used were Cori at National Energy Research Scientific Computing Center (NERSC), which is supported by the Office of Science of the US Department of Energy under contract no. DE-AC02-05CH11231, Stampede2 at the Texas Advanced Computing Center (TACC) through Extreme Science and Engineering Discovery Environment (XSEDE), which is supported by National Science Foundation (NSF) under grant no. ACI-1053575, Frontera at TACC, which is supported by NSF grant no. OAC-1818253, and Bridges-2 at the Pittsburgh Supercomputing Center (PSC), which is supported by NSF award number ACI-1928147. We thank Hyungjun Lee for technical assistance with the EPW code.

# Supplemental Material for "Prediction of High Temperature Superconductivity in *C*2/*c*-24 Solid Hydrogen"

Mehmet Dogan[1,2], Sehoon Oh[1,2,3], Marvin L. Cohen[1,2,*]

[1] Department of Physics, University of California, Berkeley, CA 94720, USA
[2] Materials Sciences Division, Lawrence Berkeley National Laboratory, Berkeley, CA 94720, USA
[*] To whom correspondence should be addressed: mlcohen@berkeley.edu

**Computational Methods**

We compute optimized crystal structures using density functional theory (DFT) in the Perdew–Burke–Ernzerhof generalized gradient approximation (PBE GGA), [1] using the QUANTUM ESPRESSO software package. We use norm-conserving pseudopotentials with a 120 Ry plane-wave energy cutoff. [2,3] We use 16×16×8, Monkhorst–Pack k-point meshes to sample the Brillouin zone for the *C2/c*-24 phase. All atomic coordinates are relaxed until the forces on all the atoms are less than $10^{-4}$ Ry/$a_0$ in all three Cartesian directions ($a_0$: Bohr radius). After each structure has been relaxed at a given pressure, a denser sampling of the Brillouin zone is examined to determine the band structure, using a k-point mesh that is twice as fine in all three reciprocal lattice directions. The Fermi surfaces are plotted using the XCrySDen package. [4]

Using density functional perturbation theory (DFPT), [5,6] we calculate the vibrational modes in the harmonic approximation for the pressures of 400, 450 and 500 GPa. Full phonon dispersions are calculated using a 4×4×2 *q*-point sampling. We apply the anharmonic corrections using the self-consistent phonon method as implemented in the ALAMODE package. [7,8] For the anharmonic corrections, we use the finite-displacement approach with a 2×2×1 real-space supercell containing 96 atoms. [9] The displacements from the equilibrium position are 0.01, 0.03, and 0.04 Å for the second, third, and fourth order terms of the interatomic force constant, respectively. We use the EPW code to implement the Wannier function-based *k*-space and *q*-space sampling for electron–phonon coupling calculations as well as isotropic Eliashberg theory calculations. [10] The Wannier interpolation is done using a fine 64×64×32 *k*-point sampling and a fine 16×16×8 *q*-point sampling, using Wannierization with H 1s orbitals as initial projections, (–30 eV, 10 eV) as the inner window and (–30 eV, 70 eV) as the outer window around the Fermi energy. To demonstrate the accuracy of the interpolation in the inner window, the Wannier-interpolated band structure along with the DFT band structure at 400 GPa is presented in **Figure S8**. The comparison is essentially the same for the other pressures.

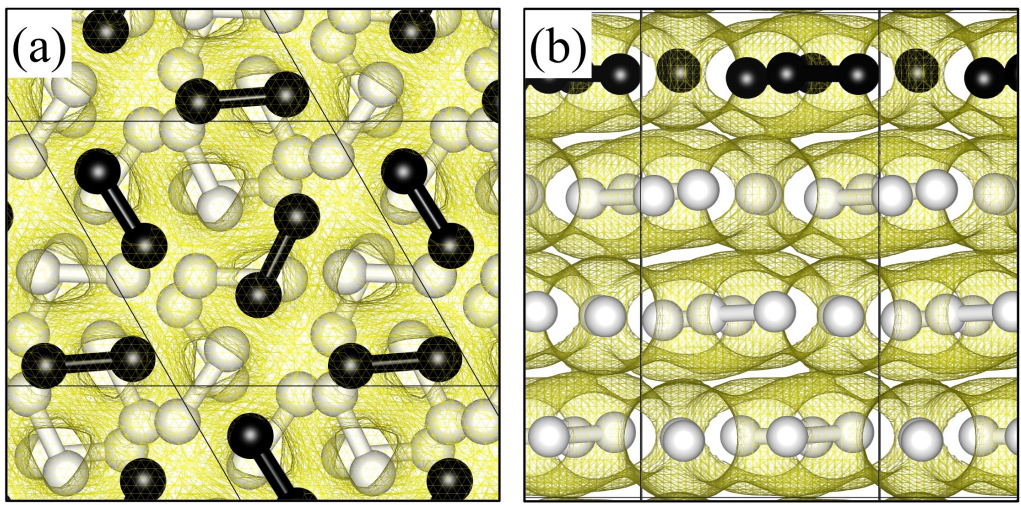

**Figure S1. Atomic structure and ELF isosurface taken at the networking value 0.276. (a-b)** The atomic structure of the *C2/c*-24 phase in the top view (a) and side view (b). The unit cell is also shown as a frame. The top layer of the unit cell is highlighted by black H atoms and bonds to ease inspection. The electron localization function (ELF) is plotted in the figures as an isosurface with value 0.276, which corresponds to the networking value (see main text for details).

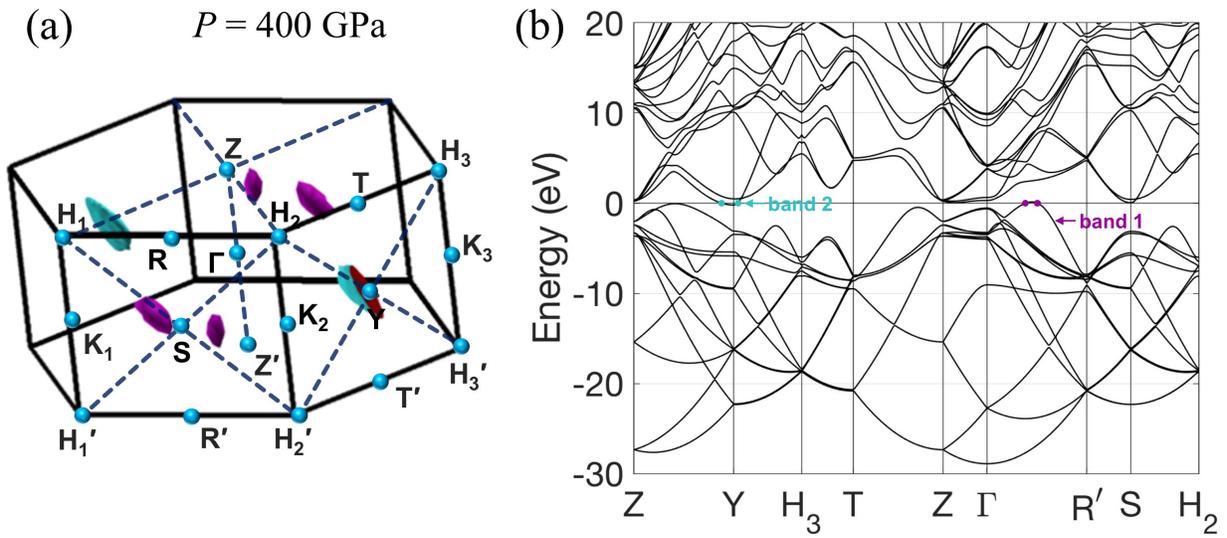

**Figure S2. Fermi surface and band structure at 400 GPa. (a)** The first Brillouin zone and the Fermi surface of the *C2/c*-24 phase of hydrogen at 400 GPa pressure, with high-symmetry points labeled. **(b)** Band structure of the *C2/c*-24 phase of hydrogen at 400 GPa pressure. Energies are relative to the Fermi energy. The bands that cross the Fermi energy are pointed out and the band crossings are labeled with filled circles. The colors of these labels match the outer surface colors of the Fermi surface sheets in (a).

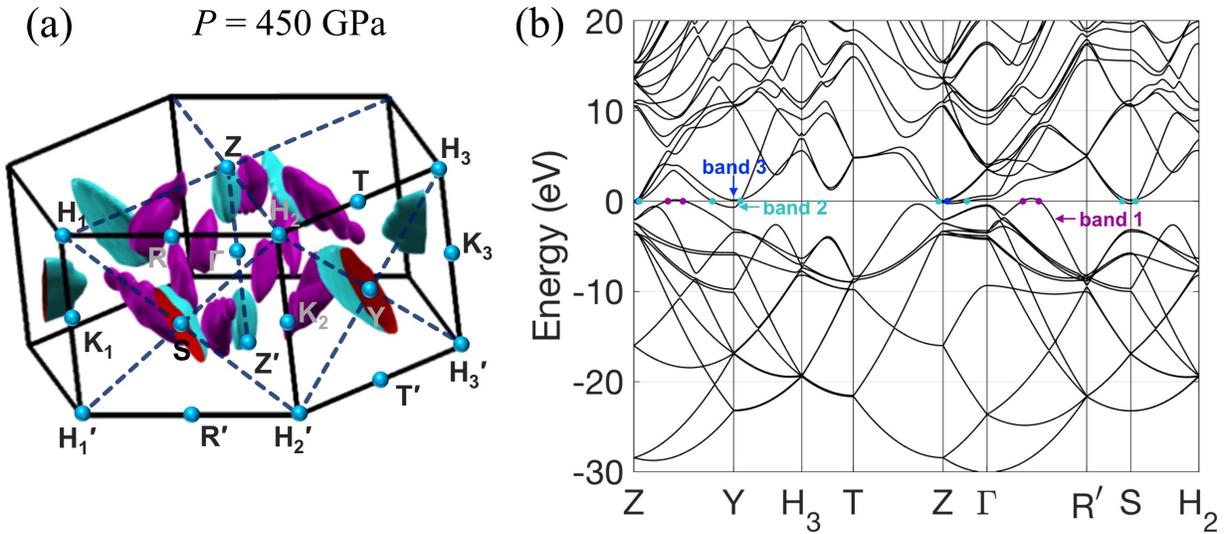

**Figure S3. Fermi surface and band structure at 450 GPa. (a)** The first Brillouin zone and the Fermi surface of the *C2/c*-24 phase of hydrogen at 450 GPa pressure, with high-symmetry points labeled. **(b)** Band structure of the *C2/c*-24 phase of hydrogen at 450 GPa pressure. Energies are relative to the Fermi energy. The bands that cross the Fermi energy are pointed out and the band crossings are labeled with filled circles. The colors of these labels match the outer surface colors of the Fermi surface sheets in (a).

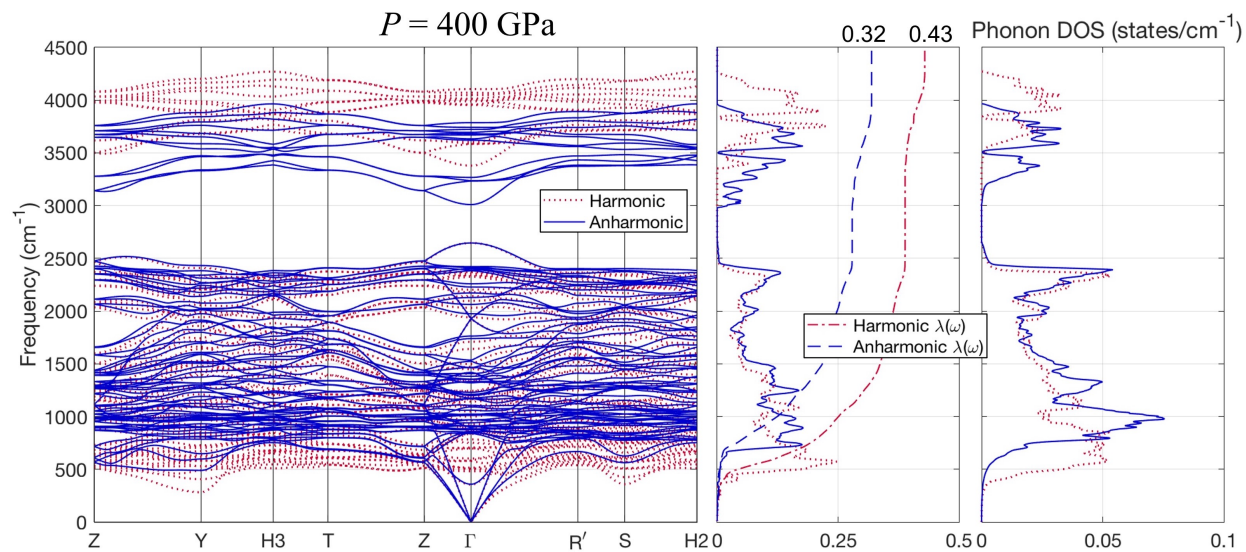

**Figure S4. Phonons and electron–phonon coupling at 400 GPa.** The phonon dispersion relations for the *C2/c*-24 phase of hydrogen at 400 GPa pressure (left panel). The harmonic and anharmonic calculations are shown by red dashed lines and blue solid lines, respectively. The Eliashberg function $\alpha^2 F$ and the electron–phonon coupling parameter $\lambda(\omega)$ (middle panel), and the phonon densities of states (right panel).

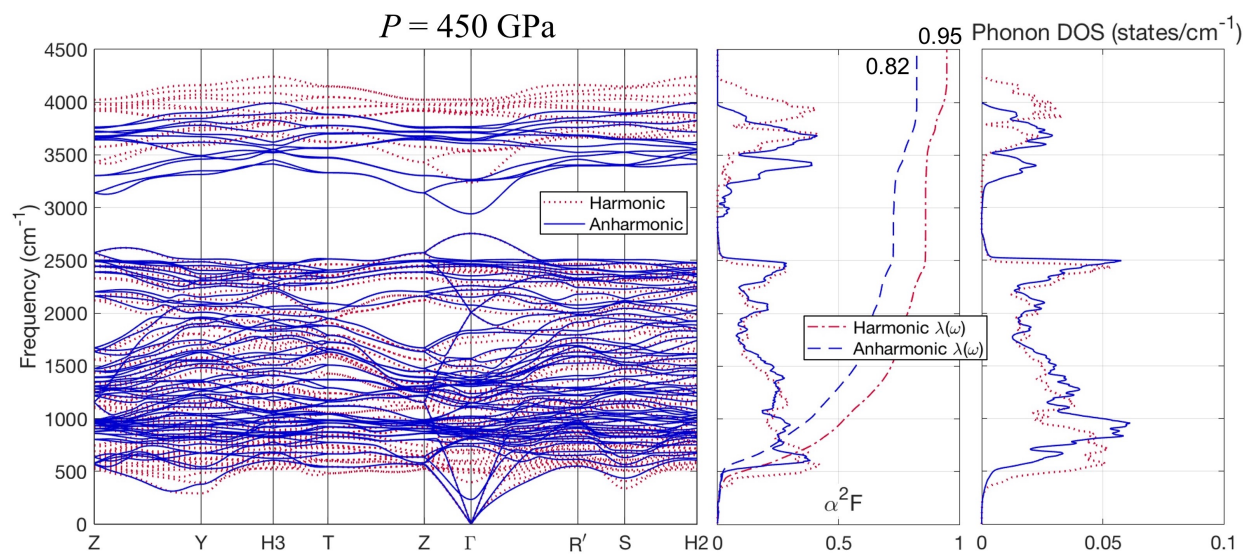

**Figure S5. Phonons and electron–phonon coupling at 450 GPa.** The phonon dispersion relations for the *C2/c*-24 phase of hydrogen at 450 GPa pressure (left panel). The harmonic and anharmonic calculations are shown by red dashed lines and blue solid lines, respectively. The Eliashberg function $\alpha^2 F$ and the electron–phonon coupling parameter $\lambda(\omega)$ (middle panel), and the phonon densities of states (right panel).

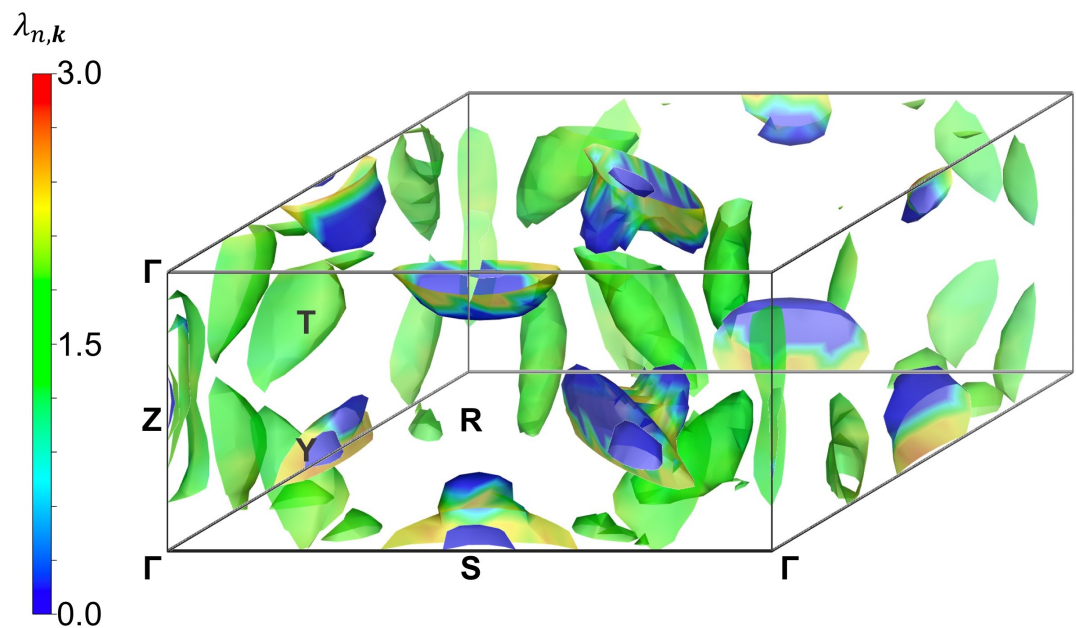

**Figure S6. Electron–phonon coupling in *k*-space for 500 GPa (harmonic).** Band and *k*-point resolved electron–phonon coupling for the *C2/c*-24 phase of hydrogen at 500 GPa in the harmonic approximation shown in the reciprocal space.

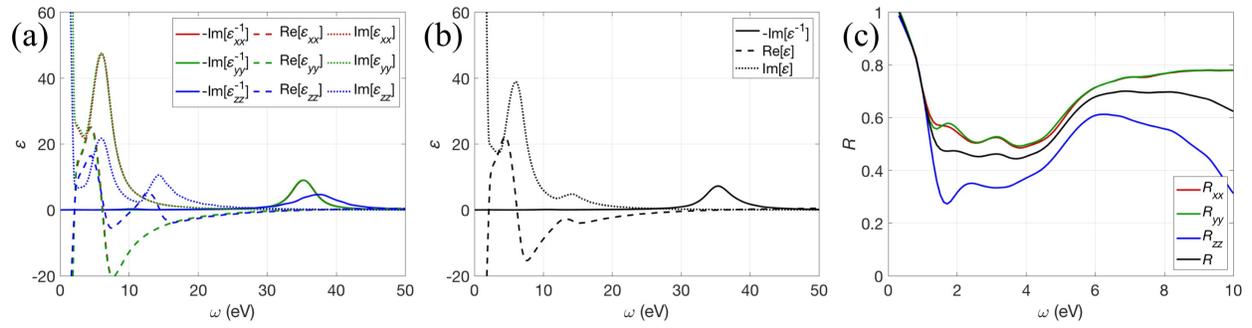

**Figure S7. Dielectric function and reflectivity at 500 GPa.** Dielectric function and reflectivity of the *C2/c*-24 phase of hydrogen at 500 GPa in the random phase approximation (RPA). **(a)** The components of $\varepsilon$ are plotted separately. **(b)** The components of $\varepsilon$ are averaged. **(c)** Reflectivity components are plotted in color and average reflectivity is plotted in black.

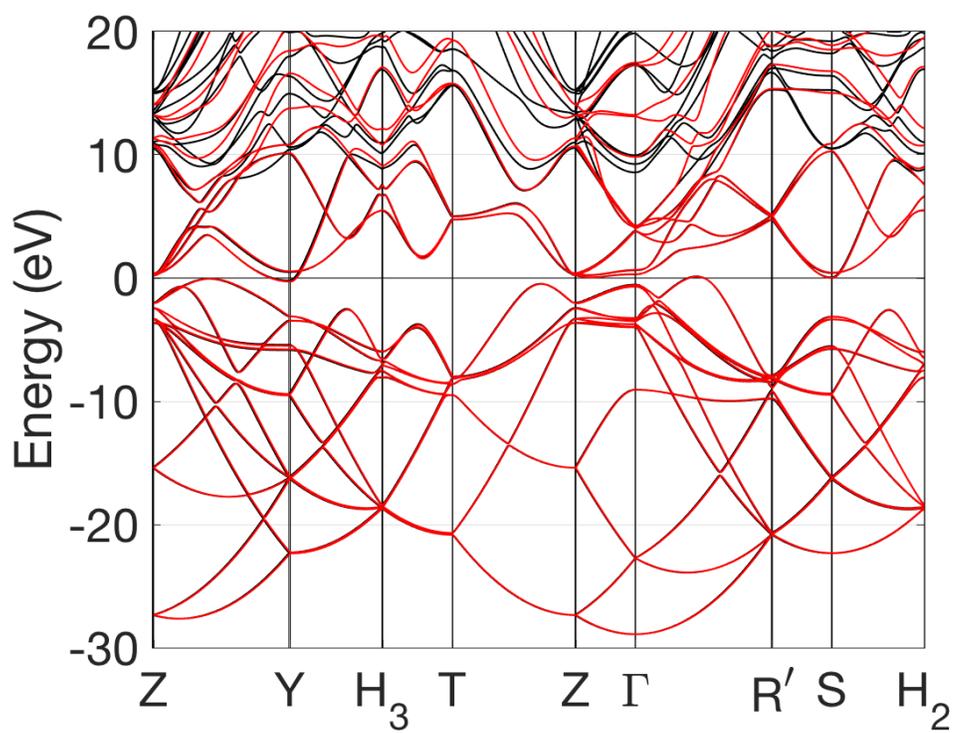

**Figure S8. Original and Wannier-interpolated band structure at 400 GPa.** The Wannier-interpolated band structure (red) overlaid on the original DFT-calculated band structure (black) at 400 GPa.